# Anisotropy of Earth's D" layer and stacking faults in MgSiO$_3$ post-perovskite

Artem R. Oganov[1*], Roman Martoňák[2], Alessandro Laio[2], Paolo Raiteri[2], Michele Parrinello[2]

[1]*Laboratory of Crystallography, Department of Materials ETH Hönggerberg, HCI G 515, Wolfgang-Pauli-Str. 10, CH-8093 Zurich, Switzerland.*

[2] *Computational Science, Department of Chemistry and Applied Biosciences, ETH Zurich, USI Campus, Via Giuseppe Buffi 13, CH-6900 Lugano, Switzerland.*

**The post-perovskite phase of (Mg,Fe)SiO$_3$ (*PPv*) is believed to be the main mineral phase of the Earth's D" layer (2700-2890 km depths). Its properties explain [1-6] numerous geophysical anomalies associated with this layer: e.g., the D" discontinuity [7], its topography [8] and seismic anisotropy [9]. Here, using a novel simulation technique, first-principles metadynamics, we identify a family of low-energy polytypic stacking-fault structures intermediate between perovskite (*Pv*) and *PPv*. Metadynamics trajectories identify plane sliding involving the formation of stacking faults as the most favourable pathway for the *Pv-PPv* phase transition, and as a likely mechanism for plastic deformation of *Pv* and *PPv*. In particular, the predicted slip planes are {010} for *Pv* (consistent with experiment [10,11]) and {110} for *PPv* (in contrast to the previously expected {010} slip planes [1-4]). Dominant slip planes define the lattice preferred orientation and elastic anisotropy of the texture. With {110} slip planes in *PPv*, we obtain a new interpretation of the shear-wave anisotropy in the D" layer, requiring a much smaller degree of lattice preferred orientation and more consistent with geophysical observations.**

The stability and properties of the *PPv* phase of (Mg,Fe)SiO$_3$ at conditions of the Earth's D" layer are extensively used to explain seismic features of this layer [1-4], to understand the observed geochemical anomalies [6] and global dynamics and evolution of the Earth [5,6,12]. The initial finding of *PPv* [1,2] was achieved with input from both experiment and theory. Here, starting from MgSiO$_3$ *Pv* and applying a new simulation technique [13], we obtain the *PPv* structure purely from first principles. This shows the potential of this simulation methodology and provides valuable new insight into the mineralogy and physics of the Earth's D" layer.

We employ the method proposed by Martoňák et al. [13,14] and based on the ideas of metadynamics [15]. In this method, one introduces an order parameter – we use the lattice vectors matrix **h** = ($h_{11}, h_{22}, h_{33}, h_{12}, h_{13}, h_{23}$) chosen in the upper triangular form. This order parameter follows discrete evolution:

$$\mathbf{h}^{t+1} = \mathbf{h}^t + \delta h \frac{\phi^t}{|\phi^t|} \quad , \tag{1}$$

where $\delta h$ is a stepping parameter, and the driving force $\phi^t = -\frac{\partial G^t}{\partial \mathbf{h}}$ is calculated from the history-dependent Gibbs potential $G^t(\mathbf{h})$ containing Gaussians added on top of the real free energy surface $G(\mathbf{h})$:

$$G^t(\mathbf{h}) = G(\mathbf{h}) + \sum_{t'<t} W e^{-\frac{|h-h^{t'}|^2}{2\delta h^2}} \quad , \tag{2}$$

where *W* is the height of the Gaussians. The derivative of the first term on the right-hand side of (2) is:

$$-\frac{\partial G}{\partial h_{ij}} = V[\mathbf{h}^{-1}(p-P)]_{ji} \quad , \tag{3}$$

where *p* and *P* are the calculated and target stress tensors, respectively.

Stress tensors are calculated from *NVT*-molecular dynamics simulations; adding Gaussians (2) and evolving **h**-matrices as described above allows one to fill the free energy wells and move the system across the lowest barrier into the domain of another structure. Thus, one finds new crystal structures and structural transformation pathways; although the latter will in general depend on the system size, precious suggestions can be inferred. To make the exploration of the free energy surface as complete as possible, it is useful to repeat simulations starting from each found structure.

We have performed classical (using a simplified interatomic potential [16] with the DL_POLY code [17]) and *ab initio* (using the VASP code [18]) simulations. While the main results discussed here were obtained *ab initio*, classical simulations were used for initial exploration of the system and for testing conditions for *ab initio* simulations (system size, run length, $\delta h$ and W parameters). *Ab initio* simulations were based on the generalized gradient approximation [19] and the all-electron PAW method [20,21]. Timestep for molecular dynamics was set to 1 fs, in classical runs we used 4 ps for equilibration and 1 ps for calculating the stress tensor; in *ab initio* calculations 0.7 ps was used for equilibration and 0.3 ps for stress tensor calculations. Simulated conditions are 200 GPa, 2000 K (classical) and 150 GPa, 1500 K (*ab initio*).

The supercells used in our calculations contained 160 atoms (4x1x2 for *PPv*, 2x2x2 for *Pv*), which is sufficiently big to encompass a large range of structures, while computationally tractable and providing clear transition paths. We used 450 eV plane-wave cutoff and the Γ-point for Brillouin zone sampling; PAW potentials had [He] core (radius 1.52 a.u.) for O, [Ne] core (1.5 a.u.) for Si, [Ne] core (2.0 a.u.) for Mg. The metadynamics parameters we used are: $\delta h = 1$ Å and $W = 38$ kJ/mol in the classical case, and 0.98 Å and $W = 32$ kJ/mol in the *ab initio* case.

Starting from *Pv* (Fig. 1a), our *ab initio* simulations first found the 3x1 structure (Fig. 1c) and then the *PPv* structure (Fig. 1d). The reverse transition pathway was found in *ab initio* simulations starting from *PPv*. Classical simulations gave all these structures, plus the 2x2 structure (Fig. 1b). Table 1 reports parameters of the Vinet equation of state [22] fitted to our *ab initio* E(V) results for these phases.

It is easy to see that these phases form a continuous family: by simple sliding of the $\{010\}_{pv}$ planes of the *Pv* structure, one can generate all the other structures. Differing only in the stacking sequence of identical layers, these structures can be described as polytypes. Polytypes had been expected [2] in *PPv* since its discovery, because its structure contains layers of $SiO_6$-octahedra and polytypism is common in layered structures. However, the polytypes found here are radically different from those that were expected: they are not based on sheets of silicate octahedra parallel to $\{010\}_{ppv}$.

Fig. 2 shows that all these polytypes become more favourable than *Pv* at sufficiently high pressure, but only the end members of this polytypic series, *Pv* and *PPv*, are thermodynamically stable at $T = 0$ K: *Pv* below 100 GPa, *PPv* above 100 GPa. Remarkably, the intermediate polytypes are only ~20-30 meV/atom higher in enthalpy around 100 GPa and could thus be easily stabilised by temperature and impurities and be present as minor phases in the D" layer.

The stacking-fault enthalpy in *PPv* at 120 GPa is only 32 meV/Å$^2$ = 513 mJ/m$^2$, a small value similar to those found in metals at ambient pressure [23]. It is common for polytypes that their typically low-energy stacking fault planes play the role of the dominant slip planes. The $\{110\}_{ppv}$ slip planes at first seem counterintuitive because they cut through the silicate sheets of the *PPv* structure, yet they are favourable. As shown by Legrand [23] on the example of hcp-metals, the product of the stacking fault energy (or enthalpy) $\gamma$ and the relevant shear elastic constant $C_s$ determines the importance of a given slip plane. This criterion works very well (even though it does not explicitly account for dislocations), because it simultaneously accounts for ease of shear and formation of energetically favourable structures during plastic deformation. In our case, if the ratio $R=\frac{\gamma_{010}C_{s,010}}{\gamma_{110}C_{s,110}}$ is greater than 1, $\{110\}_{ppv}$ slip planes should be preferred to $\{010\}_{ppv}$ slip planes. Considering different types of $\{010\}_{ppv}$ stacking faults in post-perovskite for the most stable ones (Supplementary Figure 1) we found $\gamma_{010}$ = 330 meV/ Å$^2$. Using suitably transformed elastic constants [2] we obtain $R$ = 9.5 at 120 GPa, ruling out $\{010\}_{ppv}$ slip planes in favour of $\{110\}_{ppv}$ slip.

Preferred orientation along $\{110\}_{ppv}$ explains why in diamond-anvil cell experiments on *PPv* (e.g., Ref. 1) the $\{110\}$ diffraction intensities are often vanishingly small. The fact that metadynamics could identify the most plausible slip plane in a single simulation is not surprising: the method by construction looks for the easiest non-elastic deformation mechanism and for the most energetically favourable structures along the deformation path. The mechanism of plastic deformation found here can operate also at pressures far away from the transition pressure and can be expected to be effective in analogous compounds. Indeed, using the analogy with CaTiO$_3$, Karato et al. [10] concluded that the dominant slip plane in *Pv* should be $\{010\}_{pv}$ (in the *Pbnm* setting used also in this paper); as seen in Fig.1, sliding $\{010\}_{pv}$ planes of *Pv* structure produces the *PPv* structure. The $\{010\}_{pv}$ slip was also demonstrated to be important, though not dominant, in deformation experiments of Cordier et al. [11] conducted at 25 GPa.

These $\{110\}_{ppv}$ slip planes call for a reinterpretation of the seismic anisotropy of the D" layer. Using the method of Montagner and Nataf [24], we estimated seismic anisotropy of *PPv* texture with the $\{110\}_{ppv}$ alignment; this required the elastic constants transformed to a new coordinate system: $C'_{ijkl} = \alpha_{ip}\alpha_{jq}\alpha_{kr}\alpha_{ls}C_{pqrs}$, where $\alpha$ is the transformation matrix and $C_{pqrs}$ are the elastic constants [2] in the standard setting. Convective flow in the D" layer is inclined in the regions of subduction, vertical in upwellings, and predominantly horizontal elsewhere. Due to the positive Clapeyron slope (Ref. 2, 25), *PPv* layer will be thicker in cold subduction regions, so if anisotropy of the D" layer is indeed related to *PPv* it should be more detectable in those regions. Detailed regional studies (see Ref. 8, 26, 27) indicate strong $v_{SH}/v_{SV} > 1$ anisotropy with an inclined axis in subduction regions, and variable anisotropy where the flow should be horizontal.

Orienting the $\{110\}_{ppv}$ slip planes (and the $[\bar{1}10]$ slip directions) vertically, we find that horizontally polarized shear waves propagate 4.1% faster than vertically polarized ones ($v_{SH}/v_{SV} = 1.041$). This anisotropy is larger than the previously reported value of 2.9% calculated with the assumption of horizontally located $\{010\}_{ppv}$ slip planes. With higher perfect-texture anisotropy obtained here only 33% (65% in the regions of maximum anisotropy) alignment is required to reproduce the geophysically inferred anisotropy [9]. Orienting the slip directions at some angle (subduction angle) to the vertical would explain the inclined anisotropy invariably observed [26,27] in the regions with $v_{SH}/v_{SV} > 1$; in the regions of horizontal convective flow $v_{SH}/v_{SV} < 1$ is expected. Previous interpretation based on $\{010\}_{ppv}$ slip was unable to explain inclined anisotropy and required unrealistically high degrees of lattice preferred orientation, 46% on average and 92% in maximally anisotropic regions.

Tsuchiya et al. [25] proposed a transition path from *Pv* to *PPv* based on shearing of the *Pv* structure in the $\{001\}_{pv}$ plane. We observe this mechanism in simulations performed on a small 20-atom cell. For a larger, 160-atom system, however, we see a less cooperative mechanism with elements of nucleation: shear producing locally stacking faults with fragments of the *PPv* structure. First, we observe the transition from *Pv* to the 3x1 structure on the 15-th metastep, and then to *PPv* on the 23-rd metastep. Starting from *PPv*, we observe the reverse transition to perovskite following exactly the same pathway and again involving stacking faults. Using more degrees of freedom for atomic relaxation, the transition path obtained in a larger cell is by construction energetically more favourable. Direct calculation of the enthalpy as a function of the reaction coordinate (Fig. 3) shows that this effect is very large: instead of an enthalpy maximum in the middle of the transition path we have a local minimum corresponding to the intermediate 3x1 structure. As a consequence, the activation barrier at 120 GPa drops from ~2.3 eV for the pure-shear mechanism of Ref. 25 to only 0.6 eV for our stacking-fault mediated mechanism.

Using a novel simulation technique, we have found the *Pv-PPv* transition mechanism and determined likely mechanisms of plastic deformation for both phases, involving the formation of stacking faults. Our predicted slip plane for *Pv* is consistent with experimental evidence. The predicted plastic slip of *PPv* is counterintuitive, but more consistent with geophysical observations than previous intuitive suggestions. In particular, it is now possible to explain the observed inclined character of anisotropy [26,27].

# References.

1. Murakami M., Hirose K., Kawamura K., Sata N., Ohishi Y. (2004). Post-perovskite phase transition in $MgSiO_3$. *Science* **304**, 855-858.

2. Oganov A.R., Ono S. (2004). Theoretical and experimental evidence for a post-perovskite phase of $MgSiO_3$ in Earth's D" layer. *Nature* **430**, 445-448.

3. Iitaka T., Hirose K., Kawamura K., Murakami M. (2004). The elasticity of the $MgSiO_3$ post-perovskite phase in the Earth's lowermost mantle. Nature **430**, 442-445.

4. Tsuchiya T., Tsuchiya J., Umemoto K., Wentzcovitch R.M. (2004). Elasticity of post-perovskite MgSiO$_3$. *Geophys. Res. Lett.* **31**, L14603.

5. Hernlund J.W., Thomas C., Tackley P.J. (2005). A doubling of the post-perovskite phase boundary and structure of the Earth's lowermost mantle. *Nature* **434**, 882-886.

6. Ono S. & Oganov A.R. (2005). *In situ* observations of phase transition between perovskite and CaIrO$_3$-type phase in MgSiO$_3$ and pyrolitic mantle composition. *Earth Planet. Sci. Lett.* **236**, 914-932.

7. Lay, T., Helmberger, D.V. (1983). A shear velocity discontinuity in the lower mantle. *Geophys. Res. Lett.* **10**, 63-66.

8. Lay T., Williams Q., Garnero E.J. (1998). The core-mantle boundary layer and deep Earth dynamics. *Nature* **392**, 461-468.

9. Panning, M., Romanowicz, B. (2004). Inferences on flow at the base of Earth's mantle based on seismic anisotropy. *Science* **303**, 351-353.

10. Karato S., Zhang S.Q., Wenk H.R. (1995). Superplasticity in Earth's lower mantle – evidence from seismic anisotropy and rock physics. *Science* **270**, 458-461.

11. Cordier P., Ungar T., Zsoldos L., Tichy G. (2004). Dislocation creep in MgSiO$_3$ perovskite at conditions of the Earth's uppermost lower mantle. *Nature* **428**, 837-840.

12. Nakagawa T., Tackley P.J. (2004). Effects of a perovskite-post perovskite phase change near core-mantle boundary in compressible mantle convection. *Geophys. Res. Lett.* **31**, L16611.

13. Martoňák R., Laio A., Parrinello M. (2003). Predicting crystal structures: The Parrinello-Rahman method revisited. *Phys. Rev. Lett.* **90**, 075503.

14. Martoňák R., Laio A., Bernasconi M., Ceriani C., Raiteri P., Zipoli F., Parrinello M. (2005). Simulation of structural phase transitions by metadynamics. *Z. Krist.* **220**, 489-498.

15. Laio A., Parrinello M. (2002). Escaping free-energy minima. *Proc. Natl. Acad. Sci.* **99**, 12562-12566.

16. Oganov A.R., Brodholt J.P., Price G.D. (2000). Comparative study of quasiharmonic lattice dynamics, molecular dynamics and Debye model in application to MgSiO$_3$ perovskite. *Phys. Earth Planet. Int.* 122, 277-288.

17. Smith W., Todorov I.T., Leslie M. (2005). The DL_POLY molecular dynamics package. *Z. Krist.* **220**, 563-567.

18. Kresse G. & Furthmüller J. (1996). Efficient iterative schemes for ab initio total-energy calculations using a plane wave basis set. *Phys. Rev.* **B54**, 11169-11186.

19. Perdew J.P., Burke K., Ernzerhof M. (1996). Generalized gradient approximation made simple. *Phys. Rev. Lett.* **77**, 3865-3868.

20. Blöchl P.E. (1994). Projector augmented-wave method. *Phys. Rev.* **B50**, 17953-17979.

21. Kresse G., Joubert D. (1999). From ultrasoft pseudopotentials to the projector augmented-wave method. *Phys. Rev.* **B59**, 1758-1775.

22. Vinet P., Rose J.H., Ferrante J., & Smith J.R. (1989). Universal features of the equation of state of solids. *J. Phys.: Condens. Matter* 1, 1941-1963.

23. Legrand B. (1984). Relations entre la structure èlectronique et la facilitè de glissement dans les métaux haxagonaux compacts. *Philos. Mag.* **49**, 171-184.

24. Montagner, J.-P., Nataf, H.-C. (1986). A simple method for inverting the azimuthal anisotropy of surface waves. *J. Geophys. Res.* **91**, 511-520.

25. Tsuchiya T., Tsuchiya J., Umemoto K., Wentzcovitch R.M. (2004). Phase transition in MgSiO$_3$ perovskite in the earth's lower mantle. *Earth Planet. Sci. Lett.* **224**, 241-248.

26. Garnero E.J., Maupin V., Lay T., Fouch M.J. (2004). Variable azimuthal anisotropy in Earth's lowermost mantle. *Science* **306**, 259-261.

27. Wookey J., Kendall J.-M., Rümpker G. (2005). Lowermost mantle anisotropy beneath the north Pacific from differential *S-ScS* splitting. *Geophys. J. Int.* **161**, 829-838.

## Acknowledgements.

Calculations were performed at ETH Zurich and CSCS (Manno). ARO is grateful to P. Cordier, T. Ungar, G. Ferraris, T. Balic-Zunic, E. Makovicky and C. Thomas for illuminating discussions on various aspects of this work.

Correspondence and requests for materials should be addressed to A.R. Oganov (a.oganov@mat.ethz.ch)

**FIGURE LEGEND:**

Fig. 1. MgSiO$_3$ polytypes found by metadynamics. a – *Pv* (space group *Pbnm*), d – *PPv* (*Cmcm*), b,c – newly found structures 2x2 (*Pbnm*) and 3x1 (*P2$_1$/m*), respectively. Only silicate octahedra are shown; Mg atoms are omitted for clarity. In the *PPv* structure, the previously expected slip plane is parallel to the sheets formed by silicate octahedra; the most likely slip plane identified here is shown by an arrow. Arrows also show slip planes in the other structures.

Fig. 2. Enthalpies (relative to *Pv*, per formula unit) of MgSiO$_3$ polytypes as a function of pressure. Solid line – *PPv*, dashed line – 2x2 structure, dotted line – 3x1 structure.

Fig. 3. Activation barrier for the *Pv-PPv* transition at 120 GPa. Dashed line – mechanism of Ref. 25, solid line – mechanism proposed here. Enthalpies are given per formula unit. Along the reaction coordinate, the **h**-matrix smoothly changes from the values characteristic of *Pv* to those of *PPv*. At each point in the plot atomic positions were optimised under constraint of fixed h-matrix.

**Table 1. Calculated equation of state parameters for MgSiO$_3$ polytypes. All results are given per 20 atoms; energies are relative to perovskite.**

| Phase | $E_0$, eV | $V_0$, Å$^3$ | $K_0$, GPa | $K_0'$ |
|---|---|---|---|---|
| Perovskite (*Pbnm*) | 0 | 167.997 | 230.87 | 4.125 |
| 2x2 (*Pbnm*) | 0.685 | 170.076 | 194.01 | 4.553 |
| 3x1 (*P2$_1$/m*) | 0.645 | 169.563 | 198.55 | 4.515 |
| Post-perovskite (*Cmcm*) | 0.928 | 168.161 | 201.79 | 4.498 |

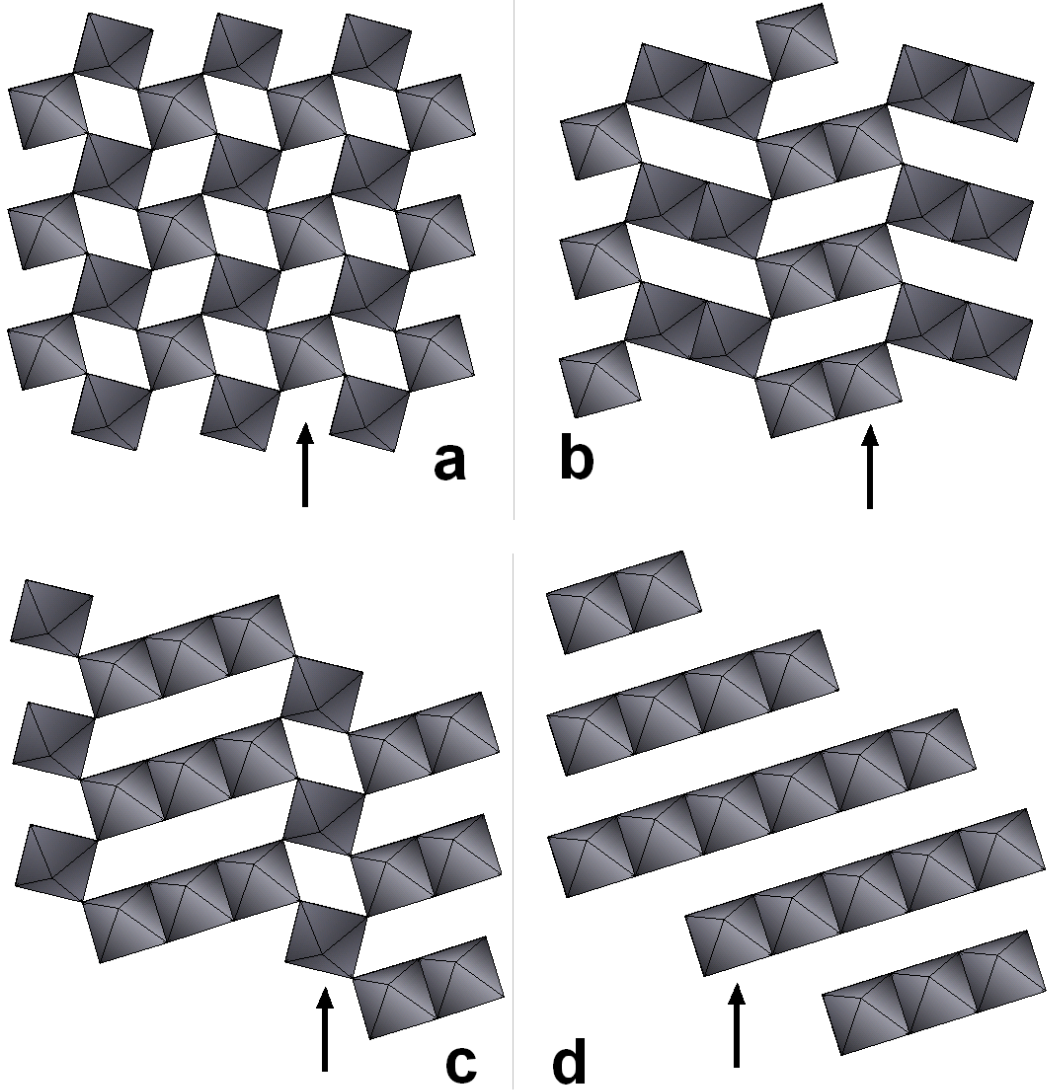

**Fig. 1. Oganov et al.**

14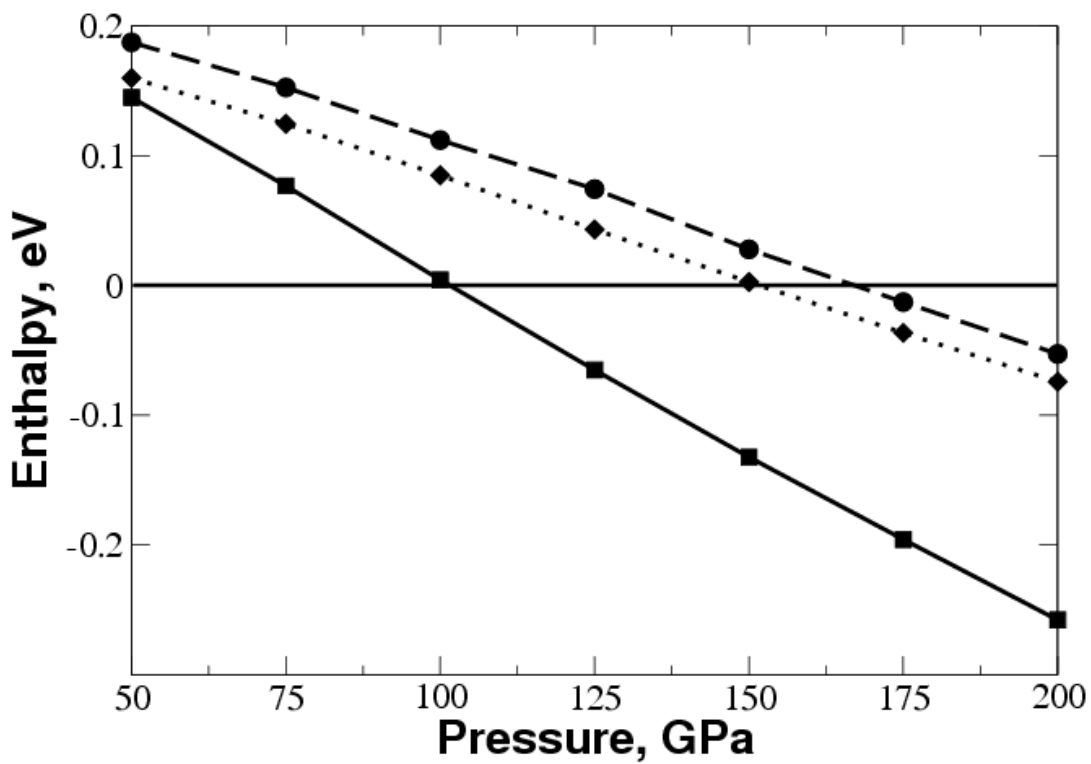

**Fig. 2. Oganov et al.**

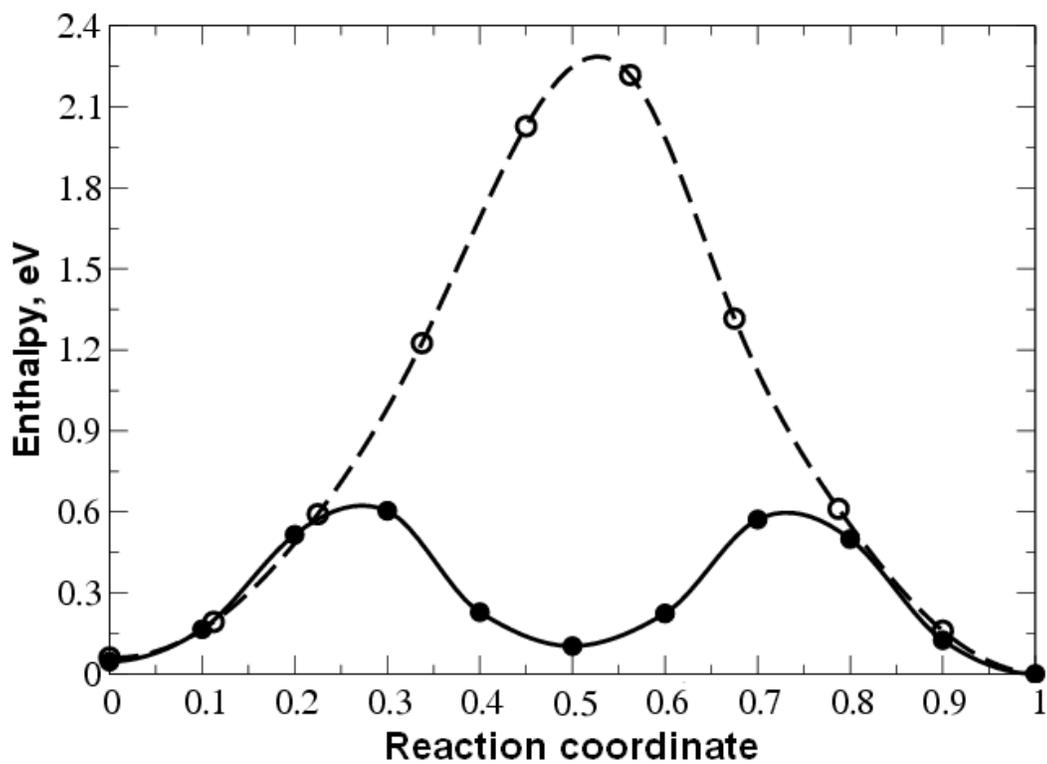

**Fig. 3. Oganov et al.**